\documentstyle[11pt]{article}
\begin{document}
%\normalsize
%\draftmode
\newcommand{\Dslash}{D\kern-0.15em\raise0.17ex\llap{/}\kern0.15em\relax}
\newcommand{\deeslash}{\partial\kern-0.15em\raise0.17ex\llap{/}
\kern0.15em\relax}
\newcommand{\Bslash}{B\kern-0.15em\raise0.17ex\llap{/}\kern0.15em\relax}
\newcommand{\Fslash}{{\cal D}\kern-0.15em\raise0.17ex\llap{/}
\kern0.15em\relax}
\newcommand{\ts}{\tilde{\sigma}}
\newcommand{\utN}{\rlap{\lower2ex\hbox{$\,\,\tilde{}$}}{N}}
\rightline{hep-th/9702171, VPI-IPPAP-97-1}
\vskip 0.20in
\centerline{\Large The Weyl Theory of Fundamental Interactions:}
\vskip 0.20in
\centerline{\Large Is CPT violated?}
\vskip 0.20in
\vskip 0.20in
\centerline{Chopin Soo ${}^\dagger$}%%%
\vskip 0.10in
\centerline{Winnipeg Institute for Theoretical Physics and}
\centerline{Physics Department, University of Winnipeg,} 
\centerline{Winnipeg, Manitoba, Canada R3C 2E9.} 
\vskip 0.10in
\centerline{and}
\vskip 0.10in
\centerline{Lay Nam Chang ${}^*$}
\vskip 0.10in
\centerline{Department of Physics and}
\centerline{Institute for Particle Physics and Astrophysics,}
\centerline{Virginia Tech,}
\centerline{Blacksburg, VA 24061-0435, U.S.A.}\vskip0.20in
\centerline{PACS number(s): 11.30.Er, 11.30.Rd, 04.62.$+$v}
\vskip 0.20in
\vskip0.20in
%\centerline{\today}
%\centerline{{\bf Abstract}}

We compare the conventional description of the interaction of matter with
the four known forces in the standard model with an alternative
Weyl description in which the chiral coupling is extended to include
gravity. The two are indistinguishable at the low energy classical
level of equations of motion, but there are subtle differences at the
quantum level when nonvanishing torsion and the Adler-Bell-Jackiw
anomaly is taken into account. The spin current and energy-momentum
of the chiral theory then contain non-Hermitian terms which are not
present in the conventional theory. In the chiral alternative, 
CPT invariance is not automatic because chirality supersedes Hermiticity but 
full Lorentz invariance holds. New fermion loop processes associated with 
the theory are discussed together with a perturbative regularization 
which explicitly maintains the chiral nature and local symmetries of the
theory.

\vfil
$^\dagger$Electronic address: soo@theory.uwinnipeg.ca \hfil
\vskip0.02in
$^*$Electronic address: laynam@vt.edu \hfil

%
%
%Introduction
%
\bigskip
\section*{I. INTRODUCTION}
\bigskip

It is well-known that in the standard model, fermions interact chirally 
with internal gauge fields and can be described by a multiplet of 
15 left-handed Weyls per generation coupled to the basic forces. 
In the absence of gravity, at least as far 
as perturbation theory is concerned, the rewriting of right-handed 
fields as left-handed ones is mere relabeling without physical consequence.
The conventional
coupling to gravity is Majorana since the Hermitian Weyl action contains
both the left- and right-handed spin connections and can
be expressed as a Majorana action in the absence of internal gauge fields.
%Yet because the fermion multiplet belongs to a complex representation of the 
%internal gauge group, it is not possible to express the complete action 
%with all four forces in pure Majorana form. 
As shown in Ref. \cite{ash}, in four dimensions, gravity
can be described using spin connections of a single-handedness. When
we bring in internal gauge fields, the difference is important, because
the fermion multiplet belongs to a complex representation of the internal gauge
group. It is not possible to express the complete action with all four
forces in pure Majorana form. Whether we write some of the fields of
one chirality as conjugate fields of the opposite chirality can produce
couplings to spin connections of different handedness. 
%
% The conventional Weyl-for-internal-gauge-group
%and Majorana-for-gravity prescription seems to be yet another subtle 
%difference between gravity and the other three forces. 
%If simplicity and economy
%are of the essence, it is certainly simpler to prescribe that all
%interactions between fermions and the gauge fields, including gravity, 
%be Weyl;
%and more economical in the  sense that Weyl coupling to gravity requires 
%only the left-handed rather than the full spin connection.
%In the presence of gravity this modification
%is not mere relabeling for a number of reasons. 

If fermions couple to only the left-handed spin connection, then what
dynamics is to be prescribed for the left-handed spin connection
without leading to spurious equations of motion for the right-handed spin 
connection? An answer that suggests itself is that when such a 
chiral Weyl theory of fermions is quantized, the one-loop
effective action generated in background curved spacetimes inevitably 
result in counterterms. These will be compatible with the  
Weyl nature  and symmetries of the theory if there are no anomalies. Thus
the lowest order curvature counterterm which is the analog of the 
Einstein-Hilbert-Palatini action for the conventional case is therefore 
the natural candidate for the classical action. As we shall see, 
this leads naturally to 
the Samuel-Jacobson-Smolin action \cite{samuel} for (anti-)self-dual
Ashtekar variables \cite{ash}. Conversely, if such an action which involves 
not the full but only the left-handed spin connection is
used, then fermions of only one chirality is allowed. 
An obvious question is whether such a theory differs in any detectable 
way from the conventional one \cite{soo, cps}.

The Hermitian Weyl action in curved spacetimes requires the 
right-handed spin connection. A Weyl action with only the 
left-handed spin connection is therefore not Hermitian in general. 
However, the chiral Weyl theory is local Lorentz invariant because in 
four dimensions, the miracle is that the relevant
$SO(3,1)$ group is isomorphic to $SO(3,C)$ and $SL(2,C)$ is the 
complexification of $SU(2)$. In fact the (anti-)self-dual Ashtekar connection
is the left-handed complexified $SU(2)$ connection when the Lorentz group
is gauged. While P and CP have been observed to be violated,
there is a bias to believe that CPT is good.
The CPT-theorem \cite{lee} says that a
local, Lorentz-invariant quantum field theory with a {\it Hermitian}
action $S$ quantized according to the spin-statistics rule
must be CPT-invariant. Under CPT, $S \rightarrow S^\dagger$. 
%Sometimes the theorem is promoted as indicating that any ``reasonable"
%quantum field theory that can account for all the basic forces and the 
%associated phenomena in four dimensions must be CPT-invariant. 
%In four dimensions there is an alternative
%description of all the  four known forces which is a completely chiral.
The Hermitian qualification is essential to CPT being good and
the chiral nature of the
standard model suggests that it is a possible candidate for a breakdown
of CPT invariance.\footnote{The violations studied here are due 
to non-Hermiticity rather than to infinite number of basic fields, 
nonlocality, strings and extended objects or a breakdown of Lorentz 
invariance.} 
As we shall see, violations of CPT are subtle and occur
in regimes which have not been rigorously tested.

In what follows, we discuss the standard model with gravitational coupling 
in the context of a truly chiral
Weyl theory \cite{cps} and detail how it differs from the conventional scheme.
We prove its consistency with respect to all the local symmetries of the
theory and provide a regularization scheme \cite{inv} 
which explicitly respects the
chirality and local symmetries and permits the computation of fermion
loops which include CPT-violating processes in the presence of torsion.
We show that the energy-momentum tensor and spin current contain 
anti-Hermitian pieces which all involve the Adler-Bell-Jackiw 
(ABJ) current \cite{ABJ}. 
The scaling behaviour of the theory also differs from the
conventional case in that the trace of the energy-momentum tensor
contains anti-Hermitian contributions involving the ABJ current and the ABJ 
anomaly.

\bigskip
\section*{II. WEYL, MAJORANA AND HERMITIAN ACTIONS}
\bigskip
  
We start with the bare chiral (Weyl) fermion action which is
\begin{eqnarray}
S^- & =&\int_M d^4x e{\overline{\Psi}}_Li
{{D\kern-0.15em\raise0.17ex\llap{/}\kern0.15em\relax}}
\Psi_{L}\cr
\nonumber\\
&=& \int_M d^4x {\tilde{\overline{\Psi}}}_L e^{1\over2}i\Dslash e^{-{1\over2}}
{\tilde\Psi}_L
\nonumber\\
\end{eqnarray}
where $i{D\kern-0.15em\raise0.17ex\llap{/}\kern0.15em\relax} =\gamma^\mu
(i\partial_\mu + W_{\mu a}T^a + \frac{i}{2}
 A_{\mu AB}{\sigma}^{AB})$, 
and $e$ denotes the determinant of the vierbein.
$W_{\mu a}$ is the internal gauge connection while $A_{\mu AB}$ is the spin 
connection.
%$P_L = \frac{1}{2}(1 - \gamma^5)$ is the left-handed projection operator. 
The convention is
\begin{equation}
\{ \gamma^{A}, \gamma^{B}\}= 2\eta^{AB},
\end{equation}
with $\eta^{AB} = {\rm diag}(-1,+1,+1,+1)$.
Lorentz indices are denoted by 
uppercase Latin indices while Greek indices are spacetime indices.

In the standard model, it is possible to 
relabel right-handed fields $\Psi_{R_i}$ as left-handed ones 
$\chi_{L_i}$  through
$\Psi_{R_i} = C_4{\overline\chi}^T_{L_i}$ 
and rewrite all Weyl fermion fields
in a multiplet $\Psi_{L}$, of 15 Weyl fermions, which belongs to a 
complex representation of the gauge group.\footnote{$i$ denotes the 
flavor/color and $C_4$ is the charge conjugation matrix in four dimensions
with $C^T_4 = C^{-1}_4 = C^\dagger_4 = -C_4$.}
The arguments and
results presented in this article are however not confined to just the
$SU(3)\times SU(2)\times U(1)$ group and 15 Weyls, but are also applicable
to Grand Unified Theories \cite{glash} coupled {\it chirally} to gravity.

Recall that if the gauge generators $T^a$ satisfy
\begin{equation}
\left[T^a, T^b \right] = if^{ab}\,_cT^c,
\end{equation}
then $(-T^a)^*$ satisfy the same Lie algebra.\footnote{We adopt the convention 
of $(T^a)^\dagger = T^a$ and real structure constants.} 
If there exists a $U$ such 
that $U^{-1}(-T^a)^* U = T^a$, then the representation is called 
real (pseudoreal) if $U$ is symmetric (antisymmetric). Otherwise, the 
representation is termed complex.

The Hermitian conjugate of the Weyl action is
\begin{equation}
(S^-)^\dagger = \int_M d^4x e\left[-i(\partial_\mu{\overline\Psi}_L)\gamma^\mu
\Psi_L +{ \overline\Psi}_L(\frac{i}{2}A_{\mu AB}\sigma^{AB} + W_{\mu a}T^a)
\gamma^\mu\Psi_L\right].
\end{equation}

Without gauge fields ($W_{\mu a}=0$), the Hermitized action is expressible in
Majorana form as
\begin{equation}
S_{Majorana}= \frac{1}{2}(S^- + (S^-)^\dagger )= \int_M d^4x e{\overline\Psi}_M
i\Dslash\Psi_M,
\end{equation}
with
\begin{eqnarray}
\Psi_M& =&\frac{1}{\sqrt 2}(\Psi_L + C_4{{\overline\Psi}_L}^T)\cr
\nonumber\\
{\overline\Psi}_M &=&\frac{1}{\sqrt 2}({\overline\Psi}_L + \Psi^T_L C_4)
\nonumber\\
\end{eqnarray}
being the Majorana spinors.
When there are internal gauge couplings, 
it is still possible to write the Hermitized Weyl action in the Majorana form
for {\it real and pseudoreal} representations since
\begin{eqnarray}
S'_{Majorana}&= &\int_M d^4x e{\overline\Psi}'_M
i\Dslash\Psi'_M\cr
\nonumber\\
&=& \frac{1}{2}(S^- + (S^-)^\dagger)+ \int_M d^4x e{\overline\Psi}_L\gamma^\mu
W_{\mu a}( U^{-1}(-T^a)^TU -T^a)\Psi_L 
\nonumber\\
\end{eqnarray}
with a different set of Majorana spinors
\begin{eqnarray}
\Psi'_M& =&\frac{1}{\sqrt 2}(\Psi_L + U^{-1} C_4{{\overline\Psi}_L}^T),\cr
\nonumber\\
{\overline\Psi}'_M &=&\frac{1}{\sqrt 2}({\overline\Psi}_L + \Psi^T_L C_4 U),
\nonumber\\
\end{eqnarray}
and the choice of $U$ for which $U(-T^a)^T U^{-1}- T^a = 0$ can be found. 
For complex representations however, the Hermitized action has a Majorana
form only for gravity couplings.

We next examine the case in which the coupling of matter
to all four known fources including gravity is 
{\it entirely chiral or Weyl} i.e.
the couplings are described precisely by $S^-[{\overline\Psi}_L, \Psi_L, 
e_{\mu A}, A^-_{\mu AB}, W_{\mu a}]$ as in Eq. (1), 
rather than its Hermitian form. 
The quantity $A^-_{\mu AB} = {1\over 2}
(-iA_{AB}+{1\over2}\epsilon_{AB}\,^{CD}A_{AB})$ is the
anti-self-dual or left-handed spin connection, and only
left-handed fermions appear in the action.\footnote{Note that
in $(S^-)^\dagger$ in Eq. (4) the coupling projects out the right-handed spin 
connection while in $S^-$ only the left-handed spin 
connection is required.  
So a Majorana fermion couples to both left- and right-handed spin connections. 
The same is true for a Weyl fermion in the Hermitized action. 
See for instance Ref. \cite{cps} for explicit
self- and anti-self-dual decompositions of the spin connection and
their couplings to two-component Weyl fermions.} 

For clarity, we shall use the explicit chiral 
representation with
\begin{eqnarray}
\gamma^{5}&=& \left(\matrix{1_{2}&0\cr 0&-1_{2}}\right), \cr
\nonumber\\
\gamma^{A}&= &\left(\matrix{ 0& i\tau^{A}\cr i{\overline \tau}^{A}&0}
\right).
\end{eqnarray}
In the above, $\tau^{a}= -{\overline \tau}^{a}$ ($a$=1,2,3) are Pauli
 matrices, and $\tau^{0} = {\overline \tau}^{0} = -1_{2}$. 
In the chiral representation,
\begin{equation}
\frac{i}{2}A_{\mu AB}\sigma^{AB}P_L 
= \left[\matrix{0 & 0\cr 0 & A^-_{\mu a}\frac{\tau^a}{2}}\right].
\end {equation}
Note that 
\begin{eqnarray}
A^-_{\mu a}\frac{\tau^a}{2} &=& \left[iA_{\mu 0a}- 
\frac{1}{2}\epsilon_{0a}\,^{bc}A_{\mu bc}\right]\frac{\tau^a}{2} \cr
\nonumber\\
&=& -\frac{i}{4}A^-_{\mu AB}{\overline\tau}^A\tau^B
\end{eqnarray}
is also precisely the Ashtekar connection in the (anti-)self-dual 
formulation of gravity in four dimensions \cite{ash, samuel}. 
In this regard, since right-handed spin connections do not appear at the 
fundamental level, couplings only to left-handed Weyl fermions 
are allowed \cite{cps, ART}.
 
In what follows, we shall suppose
as in first order formulations, that
the spin connection and vierbein are independent and that the 
torsion is not necessarily zero. Before isolating the pure and imaginary
parts of the action, let us first
summarize some basic relations involving torsion.

A GL(4,R) connection $\Gamma^\alpha_{\mu\nu}$ may be introduced through
\begin{equation}
\nabla_\mu E^{\nu A} =\partial_\mu E^{\nu A} 
+ \Gamma^\nu_{\alpha\mu}E^{\alpha A}
+A_{\mu AB}E^{\nu B} =0 ,
\end{equation}
and
\begin{equation}
\nabla_\mu e_\nu\,^A=\partial_\mu e_{\nu}\,^A - 
\Gamma^\alpha_{\nu\mu}e_{\alpha}\,^A
+A_{\mu AB}e_{\nu}\,^B =0 .
\end{equation}
The covariant and contravariant metrics $g_{\mu\nu} = e_\mu\,^A e_{\nu A}$ and 
$g^{\mu\nu} = E^{\mu A}E^\nu\,_{A}$
therefore satisfy the metricity condition (which preserves lengths under
parallel transport) with respect to the connection $\Gamma^{\alpha}_{\mu\nu}$
i.e.
\begin{equation}
\nabla_\mu g_{\alpha\beta}=\nabla_\mu g^{\alpha\beta} =0.
\end{equation}

The connection $\Gamma^\alpha_{\mu\nu}$ is not necessarily torsionless or
symmetric. In fact,
\begin{equation}
\Gamma^\alpha_{[\mu\nu]}={1\over2}(\Gamma^\alpha_{\mu\nu}- 
\Gamma^\alpha_{\nu\mu})
\end{equation}
is related to the torsion
\begin{equation}
T_A = {1\over2} T_{A\mu\nu}dx^\mu\wedge dx^\nu = de_A + A_{AB}\wedge e^B
\end{equation}
by 
\begin{equation}
T_{A\mu\nu} = 2\Gamma^\alpha_{[\nu\mu]}e_{\alpha A}.
\end{equation}

In the presence of torsion, the Dirac matrices 
$\gamma^\mu = E^{\mu}_A\gamma^A$ satisfy
\begin{equation}
\partial_\mu \gamma^\mu +(\partial_\mu \ln e)\gamma^\mu +
{1\over2}A_{\mu AB}[\sigma^{AB},\gamma^\mu] = 2\Gamma^\nu_{[\nu\mu]}
\gamma^\mu
\end{equation}
with $\partial_\mu\ln e = \Gamma^\nu_{\nu\mu}$ and
${\sigma}^{AB}={1\over 4}[\gamma^A, \gamma^B]$. 
However, if the torsion vanishes, then
\begin{equation}
D_\mu e{\gamma^\mu} f = e{\gamma^\mu}{D_\mu}f.
\end{equation}

In the quantum theory,
to be compatible with the diffeomorphism-invariant 
measure \cite{Fujienergy}
\begin{equation}
\prod_{x}D[{\overline\Psi}_{L}(x)e^{1\over2}(x)]
D[e^{1\over2}(x)\Psi_{L}(x)],
\end{equation}
in curved spacetimes, 
we shall chose densitized variables defined by
\begin{eqnarray}
{\tilde{\overline\Psi}}_{L} &\equiv& {\overline\Psi}_{L} e^{1\over2},\cr
\nonumber\\
{\tilde\Psi}_{L} &\equiv& e^{1\over2}{\Psi}_{L}.
\nonumber\\ 
\end{eqnarray}
This accounts for the second line of Eq. (1) in the Weyl action.

We may  now decompose the Weyl action into explictly Hermitian 
and anti-Hermitian pieces by writing
\begin{equation}
S^- = \frac{1}{2}( S^- + (S^-)^\dagger) + \frac{1}{2}(S^- - (S^-)^\dagger),
\end{equation}
with the pure imaginary piece $iIm(S^-)$ given by
\begin{equation}
\frac{1}{2}(S^- - (S^-)^\dagger) = \int_M \left(d^4x \frac{i}{2}
\partial_\mu({\tilde {\overline\Psi}}_L\gamma^\mu{\tilde\Psi}_L)
-\frac{i}{4}{\tilde {\overline\Psi}}_L\gamma^A{\tilde\Psi}_Le^{-1}
\epsilon_{ABCD}T^B\wedge e^C \wedge e^D \right).
\end{equation}
We may note that in the presence of torsion, the second term is 
{\it not a total divergence} and 
therefore gives rise to anti-Hermitian contributions to the 
energy-momentum and spin current, but not the internal gauge current.
We shall return to these later on in Section VII.

Recalling that $T_{A\mu\nu} = 2\Gamma^\alpha_{[\nu\mu]}e_{\alpha A}$, and
defining $\Gamma^{\nu}_{[\nu\mu]}  \equiv B_\mu$,
the imaginary part of the Weyl action is in fact
\begin{eqnarray}
iIm(S^-) &=& \frac{1}{2}(S^- - (S^-)^\dagger) \cr
\nonumber\\
&=& \frac{i}{2}\int_M d^4x \left[
\partial_\mu({\tilde {\overline\Psi}}_L\gamma^\mu{\tilde\Psi}_L)
-2B_\mu{\tilde {\overline\Psi}}_L\gamma^\mu{\tilde\Psi}_L \right].
\nonumber\\
\end{eqnarray}

It is interesting to realise that
$B_\mu$ actually transforms as an Abelian gauge potential \cite{nay}
\begin{equation}
B_\mu \rightarrow B_\mu + {3\over2}\partial_\mu\alpha
\end{equation}
under {\it local} scaling 
\begin{eqnarray}
e_{\mu A} &\rightarrow& \exp(\alpha(x))e_{\mu A}, \cr
\nonumber\\
E^A\,_{\mu} &\rightarrow& \exp(-\alpha(x))E^A\,_{\mu}, \cr
\nonumber\\
\Psi_L &\rightarrow& \exp(-\frac{3}{2}\alpha(x))\Psi_L, \qquad
{\overline\Psi}_L \rightarrow \exp(-\frac{3}{2}\alpha(x)){\overline\Psi}_L,\cr
\nonumber\\
A_{\mu AB} &\rightarrow& A_{\mu AB}, \qquad W_{\mu a}\rightarrow W_{\mu a}.
\nonumber\\
\end{eqnarray}
For the densitized variables, under scaling
\begin{equation}
{\tilde\Psi}_L \rightarrow 
\exp(\frac{1}{2}\alpha(x)){\tilde\Psi}_L, \qquad
{\tilde{\overline\Psi}}_L \rightarrow 
\exp(\frac{1}{2}\alpha(x)){\tilde{\overline\Psi}}_L.
\end{equation}
However, it is important to observe that the gauge group parametrized by
$\exp(\frac{1}{2}\alpha(x))$ is noncompact rather than $U(1)$.
This accounts for an imaginary term in the change of the action under 
local scaling which we shall address in Section VI.
Moreover, here we do not assume that the underlying theory has this {\it local}
scaling symmetry but rather possesses a global invariance 
under scaling. 
In fact, the Weyl fermion action
is not invariant under such a local scaling but its
Hermitized version is.
As we shall see, this global symmetry is also broken in the quantum theory by
regularization and is related to the conformal anomaly of the theory.

%begin newstuff
The ordinary divergence of the (densitized)\footnote{All currents in this 
article are densitized tensors of weight 1.} ABJ or axial current 
\begin{equation}
J^\mu_5 = -{\tilde{\overline\Psi}}_L{\gamma^\mu}{\tilde\Psi}_L =-J^\mu
\end{equation}
is related to its
covariant divergence and torsion through
\begin{eqnarray}
\partial_\mu({\tilde{\overline\Psi}}_L\gamma^\mu{\tilde\Psi}_L)
&=&e[\partial_\mu({\overline\Psi}_L\gamma^\mu\Psi_L) +\Gamma^\nu_{\nu\mu}
{\overline\Psi}_L\gamma^\mu\Psi_L]\cr
\nonumber\\
&=& e\nabla_\mu({\overline\Psi}_L{\gamma^\mu}\Psi_L) 
+ 2e\Gamma^\nu_{[\nu\mu]}{\overline\Psi}_L\gamma^\mu\Psi_L.
\nonumber\\
\end{eqnarray}

Therefore, in terms of undensitized variables,
\begin{equation}
iIm(S^-)=\frac{i}{2}\int_M d^4x e\nabla_\mu({\overline\Psi}_L 
\gamma^\mu \Psi_L).
\end{equation}

We may note from Eq.(24) that the Abelian conformal gauge field
is actually coupled to the
fermion current $J^\mu$ (hence the ABJ or axial current) 
which acts as the source for $B_\mu$ when there is torsion. 
%However $B_\mu$ here is not an independent field but is determined entirely 
%by $e_A$ and $A_{AB}$.

\bigskip
\section*{III. EQUATIONS OF MOTION}
\bigskip

  At first thought, we might suspect that replacing the Hermitized Weyl
action by the non-Hermitized version would lead to a different
Weyl equation and hence the theories would be inequivalent even classically 
when there is nonvanishing torsion. 
This might be expected because the two actions
are not even related by just a boundary term when there is torsion.
For instance, variation with respect to ${\overline\Psi}_L$ for
the Hermitized Weyl action ${1\over2}(S^- + S(^-)^\dagger)$ yields
the equation of motion (see also Ref. \cite{nay})
\begin{equation}
ie\gamma^\mu(D_\mu + B_\mu)\Psi_L =0
\end{equation}
if we take into account the identity (18), whereas the Weyl action $S^-$
gives
\begin{equation}
ie\gamma^\mu D_\mu\Psi_L =0.
\end{equation}

However, the dynamics of the gravitational fields 
has so far not been taken into account. 
It is important to remember that in the Hermitized action, fermions
couple to both left- {\it and right-handed} spin connection whereas
in the Weyl action, only the left-handed spin connection is involved.
In conventional theory, we introduce the Einstein-Hilbert-Palatini
action which, together with the Hermitian fermion action,
determine the classical dynamics of the vierbein and the full spin 
connection of both chiralities (or both dualities).
For the non-Hermitized Weyl action we must therefore introduce the 
dynamics of gravitation in such a way that it involves only the left-handed or 
anti-self-dual spin connection. The simplest candidate for the gravitational 
part of such a theory is the Samuel-Jacobson-Smolin action \cite{samuel}. 
Such an action is at least a reasonable 
candidate for low energy and classical gravitational dynamics. 
In fact, it can be argued that in any case the lowest order one-loop 
counterterm in the effective action when this non-Hermitized chiral fermion
theory is quantized in curved background will be the 
anti-self-dual Samuel-Jacobson-Smolin action with cosmological constant.
We shall next consider the theory specified by Eq. (1) augmented by
this gravitational term.

What is remarkable is that the combined {\it chiral} Weyl {\it and}
gravitational action gives precisely the same equations of motion 
for the vierbein, spin connection and fermions as the sum of the
Einstein-Hilbert-Palatini and Hermitian Weyl actions. This is true
despite the fact that these two versions of fundamental interactions
are not even canonically related without further restrictions
when there is nonvanishing torsion.
To see this we may use the identity\footnote{See for instance Ref. \cite{cps}.
In comparing with Ref. \cite{cps}, note that
$\overline\Psi_L\gamma^A\Psi_L = \phi^\dagger_L\tau^A\phi_L$ 
where in terms of  left-handed two-component Weyl fermions,
${\Psi_L} = \left[\matrix{0\cr \phi_L}\right].$
The total action in Ref. \cite{cps} differs from the one used here 
by an overall sign.} 
\begin{eqnarray}
S^- + S^-_{SJS} &=& 
-{1\over{16\pi G}}\int_M e^A\wedge e^B\wedge *F_{AB}
+{{2\lambda}\over{16\pi G}}\int_M (*1) 
+{1\over2}(S^- + (S^-)^\dagger)\cr
\nonumber\\
&& +{i\over2} \int_M d\left\{{1\over{3!}}(\epsilon_{ABCD}{\overline\Psi}_L
\gamma^A\Psi_L e^B\wedge e^C \wedge e^D) 
- {1\over{8\pi G}}e^A\wedge T_A
\right\}\cr
\nonumber\\
&&+{i\over{16\pi G}}\int_M \Theta_A\wedge \Theta^A
\nonumber\\
\end{eqnarray}
where 
\begin{equation}
\Theta_A = T_A + (2\pi G)\epsilon_{ABCD}{\overline\Psi}_L\gamma^B\Psi_L
e^C \wedge e^D,
\end{equation}
and $*$ denotes the Hodge duality operator.
The Samuel-Jacobson-Smolin action\footnote{The cosmological
constant term is included here for completeness.} which 
contains only 
$A^-_{AB} = {1\over2}(-iA_{AB}+{1\over2}\epsilon_{AC}\,^{CD}A_{CD})$
rather than the full spin connection $A_{AB}$ and the vierbein is
\begin{eqnarray}
S^-_{SJS} &=& -{i\over{8\pi G}} \int_M \Sigma^{-AB}\wedge F^-_{AB}
+{{i\lambda}\over{3(16\pi G)}}\int_M  \Sigma^{-AB}\wedge \Sigma^-_{AB} \cr
\nonumber\\
&=&  -{1\over{16\pi G}}\int_M \left\{ e^A\wedge e^B\wedge *F_{AB}- 
{2\lambda}(*1) \right\} -
{i\over{16\pi G}}\int_M \left\{ d(e^A\wedge T_A) - T_A\wedge T^A\right\}
\nonumber\\
\end{eqnarray}
with the anti-self-dual two-forms
\begin{eqnarray}
F^-_{AB}&=&d A^-_{AB} + A^-_{AC}\wedge A^{-C}\,_{B}\cr
\nonumber\\
&=& {1\over2}(-iF_{AB} +{1\over2}\epsilon_{AB}\,^{CD}F_{CD}),\cr
\nonumber\\
\Sigma^-_{AB} &=& {1\over2}(-ie_A\wedge e_B +{1\over2}
\epsilon_{AB}\,^{CD}e_C \wedge e_D).
\nonumber\\
\end{eqnarray}
$F_{AB}$ is the curvature of the full spin connection.

In Eq. (33), we observe that the first line, which is the 
Einstein-Hilbert-Palatini action plus the Hermitian fermion action,
is the real part of the combined chiral action, while the imaginary part
consists of a total divergence or boundary term and a term involving
$\Theta$-squared. The equations of motion are such that when their
real parts are satisfied, the imaginary parts give rise to no spurious
or extra equations of motion. More precisely, on varying with respect to
$A^-_{AB}$, we obtain
\begin{equation}
D^-\Sigma^-_{AB} ={{4\pi G}\over{3!}}{\overline\Psi}_L
\gamma^C \sigma^{AB}\Psi_L \epsilon_{CDFG}e^D\wedge e^F\wedge e^G,
\end{equation}
with unique solution 
\begin{equation}
A^-_{AB} ={1\over2}(-i\omega_{AB}+{1\over2}\epsilon_{AB}\,^{CD}\omega_{CD})
-(2\pi G)\left\{J_{[A}e_{B]} 
+{i\over2}\epsilon_{AB}\,^{CD}J_{[C}e_{D]}\right\},
\end{equation}
where $\omega_{AB}$ is the torsionless $(de_A +\omega_{AB}\wedge e^B =0)$
 spin connection and $J_A = {\overline\Psi}_L\gamma_A\Psi_L$.
Using the usual reality conditions that $\omega_{AB}, e_A$ and 
$J_A $ are real,
we equate the real and imaginary parts of Eq. (38), and deduce that on-shell,
$\Theta_A =0$. Consequently, the nonboundary imaginary part of the action 
in Eq. (33) yields equations of motion
which are automatically satisfied on-shell due to its dependence on
$\Theta$-squared. Variations with respect to
the vierbein and fermions must therefore produce the same set of equations 
of motion as from the Einstein-Hilbert-Palatini plus Hermitian Weyl action.

In particular, note that $\Theta_A=0$ leads to
\begin{equation}
T_A= -(2\pi G)\epsilon_{ABCD}{\overline\Psi}_L\gamma^B\Psi_Le^C\wedge e^D,
\end{equation}
and thus
\begin{equation}
B_\mu = - T_{A\nu\mu}E^{\nu A}
\end{equation}
vanishes when the on-shell value of torsion in Eq.(39) is substituted. 
This compensates
for the discrepancies between Eq.(31) and Eq.(32) before the dynamics of the
gravitational fields have been taken into account.

We may further work out from Eq. (33) that the complete imaginary part 
of the total action is
\begin{equation}
Im(S^- + S^-_{SJS}) = \frac{i}{2}\int_M (\partial_\mu J^\mu -2B_\mu J^\mu)
+ {i\over{16\pi G}}\int_M e^A\wedge e^B\wedge F_{AB}
\end{equation}
The first two terms come from the fermionic Weyl action while the final 
term is the dual of the Einstein-Hilbert-Palatini action which is present in
the Samuel-Jacobson-Smolin action.

\bigskip
\section*{IV. REGULARIZATION}
\bigskip

In order to define the quantum field theory of fermions in background 
curved spacetimes, it is necessary to regularize divergent fermion loops.
However, because chirality of the fermions
and (anti-)self-duality of the spin connection are of the essence here, 
it is desirable to have a regularization which explicitly maintains 
not just the local symmetries but also the chirality and (anti-)self-dual
coupling to gravity. 
To this end, various generic techniques such
dimensional and zeta function regularizations fall short.
Indeed to preserve explicit Lorentz and diffeomorphism invariance, spectator 
fields which do not couple to gravity are also not allowed.
Furthermore, with only a single left-handed multiplet, Lorentz-invariant 
mass terms are  Majorana in nature. The simple form of 
$\Psi^T_L C_4 \Psi_L+ H. c. $ is not invariant under internal symmetry 
transformations.
With real representations, an invariant mass term 
$m\Psi^T_L U C_4 \Psi_L + H. c.$ can be 
constructed from a single multiplet. 
However for complex representations, a gauge and Lorentz 
invariant mass term cannot be made out of a single multiplet. 
Thus in the standard model there can be no bare masses. 
This poses a challenge for invariant 
Pauli-Villars-Gupta regularization \cite{Pauli}, even 
though the chiral fermions belong to an anomaly-free representation.  
A generalization of the method of Frolov and Slavnov using an 
infinite tower of anticommuting and commuting regulators 
which are doubled in internal space \cite{FS} has been shown to be a suitable 
regularization scheme \cite{inv}. 
This proposed regularization retains the chiral (Weyl)
nature of the theory even to the extent of the coupling of matter to
gravity in that no right-handed fields and right-handed spin 
connections are introduced.

In this Section, we recount the scheme\footnote{ A more 
thorough discussion can be found in
Ref. \cite{inv}.} and illustrate how it fits into the theory. 
The perturbative theory
of the chiral fermion determinant is then defined through this regularization.
A nonperturbative definition
may also be possible if the ``overlap" formulation for the chiral
fermion determinant \cite{nar} is suitably extended to curved spacetimes
in such a manner that no right-handed spin connection and regulator fermions
are introduced. In this regard, we do not advocate that
doubling in internal gauge space is equivalent to doubling
in external space through the introduction of right-handed regulator fermion 
fields \cite{Okuyama}. While this may be true perturbatively for flat 
spacetime, to maintian local Lorentz invariance,
the introduction of 
right-handed fermions must be accompanied by that of right-handed spin
connections in curved spacetimes. The chiral coupling to gravity is thereby
disturbed by such a regularization.

Doubling in internal gauge group space to form invariant masses
for Pauli-Villars-Gupta regularization is achieved by
including fermions which transform
according to the $(-T^a)^*$ representation. An invariant mass term can be
formed because under
\begin{equation}
\Psi^-_{L_r} \rightarrow e^{i\alpha_a T^a}\Psi^-_{L_r}, \qquad
\Psi^+_{L_r} \rightarrow e^{i\alpha_a(-T^a)^*}\Psi^+_{L_r},
\end{equation}
the combination
$\left[(\Psi^+_{L_r})^TC_4 \Psi^-_{L_r} + (\Psi^-_{L_r})^TC_4 \Psi^+_{L_r} 
 + H.c.\right]$ is 
invariant under internal gauge and Lorentz transformations.
Introducing in the enlarged space the quantities
\begin{equation}
{\cal T}^a \equiv \left(\matrix{(-T^a)^* &0\cr 0& T^a}\right),\qquad 
\sigma^1 \equiv \left(\matrix{0 & 1_d\cr 1_d &0}\right),\qquad
\sigma^3 \equiv \left(\matrix{1_d & 0\cr 0 &-1_d}\right),
\end{equation}
where $d$ denotes the number of Weyls in the bare action, 
the original multiplet is projected as
\begin{equation}
\left[\matrix{0\cr \Psi_{L}}\right]={1\over2}(1_{2d}-\sigma^3)\Psi_{L_0}.
\end{equation}
 The mass terms for the regulator fermions, 
\begin{equation}
\Psi_{L_r} = \left[\matrix{\Psi^+_{L_r} \cr \Psi^-_{L_r}}\right],
\end{equation}
can be written as $m_r(\Psi^T_{L_r}\sigma^1C_4\Psi_{L_r} + H. c.).$ 
The doubled regulator fermion multiplets are to be coupled to the 
2$d$-dimensional representation of the gauge connection, $W_{\mu a}{\cal T}^a$.

The $\Psi_{L_r}$ fields are assumed to be anticommuting. Commuting doubled 
regulator fields $\Phi_{L_s}$ are introduced in a similar manner. 
These have mass terms
\begin{equation}
m_s\Phi^T_{L_s}(-i\sigma^2)C_4\Phi_{L_s}= m_s\left[-(\Phi^+_{L_s})^T
C_4\Phi^-_{L_s} +  (\Phi^-_{L_s})^TC_4\Phi^+_{L_s}\right],
\end{equation}
with
\begin{equation}
-i\sigma^2 \equiv \left(\matrix{0 & -1_d\cr 1_d & 0}\right)= \sigma^1\sigma^3.
\end{equation}
These invariant mass terms for the doubled anticommuting and 
commuting fields exist, because for the ${\cal T}^a$ representation, there
is a symmetric ($\sigma^1$) and an antisymmetric $(-i\sigma^2)$ matrix 
which satisfy
\begin{equation}
(\sigma^1){\cal T}^a (\sigma^1)^{-1}=(-i\sigma^2){\cal T}^a (-i\sigma^2)^{-1} 
= (-{\cal T}^a)^* .
\end{equation}
Note that all the fields are {\it left-handed}.

The total regularized action which is explicitly gauge and Lorentz and, also
diffeomorphism invariant 
is taken to be\footnote{We also allow all the fields to transform
under general coordinate transformations. Here, we regularize only
fermion loops in background fields, and do not address the question of 
the regularization of the gauge and gravitational fields. 
Gauge propagators may be regularized by other methods.
Full quantum gravity effects are beyond the scope of this paper.}
\begin{eqnarray}
{\cal S}_{F_{reg}}&=&{\int}d^4xe[{\sum_{r=0,2,...}}\{{\overline\Psi}_{L_r} 
i{{D\kern-0.15em\raise0.17ex\llap{/}\kern0.15em\relax}}\Psi_{L_r} 
+ {1\over 2}m_r(\Psi^T_{L_r}\sigma^1C_4\Psi_{L_r} 
+ {\overline\Psi}_{L_r}\sigma^1C^\dagger_4{\overline \Psi}^T_{L_r})\}\cr
&-&{\sum_{s=1,3,...}}\{{{\overline\Phi}_L}_s\sigma^3i
{{D\kern-0.15em\raise0.17ex\llap{/}\kern0.15em\relax}}\Phi_{L_s} + 
{1\over 2}m_s(\Phi^T_{L_s}\sigma^1\sigma^3C_4\Phi_{L_s} + 
{\overline\Phi}_{L_s}C^\dagger_4\sigma^3\sigma^1{\overline\Phi}^T_{L_s})\}].
\nonumber\\
\end{eqnarray}

The sums are over all even natural numbers for the anticommuting fields and 
over all odd natural numbers for the commuting fields. The usefulness of this 
convention will become apparent (see for instance Eq. (54).) 
With the exception of
\begin{equation} 
\Psi_{L_0} = {1\over 2}(1-\sigma^3)\Psi_{L_0}=
\left[\matrix{0\cr \Psi_{L}}\right] ,
\end{equation} 
which is the original and undoubled chiral {\it massless} 
($m_0 = 0$) fermion multiplet,
all other anticommuting $\Psi_{L_r}$ and commuting $\Phi_{L_s}$ multiplets 
are generalized Pauli-Villars-Gupta regulator fields, doubled in 
internal space, and endowed with Majorana masses, 
which we take for definiteness
to satisfy $m_n = n\Lambda$. Due to the fact that all the multiplets 
are left-handed, 
there are no couplings to the right-handed spin connection which does 
not need to be introduced for the Weyl action.

As detailed in Ref. \cite{inv}, 
the original gauge current coupled to $W_{\mu a}$ which is
\begin{equation}
J^{\mu a} =
\frac{\delta {\cal S}_F}{\delta W_{\mu a}} =
{\tilde{\overline \Psi}}_{L_0}\gamma^\mu{\cal T}^a
\frac{1}{2}(1-\sigma^3){\tilde\Psi}_{L_0},
\end{equation}
is modified by the regulators to 
\begin{eqnarray}
J^{\mu a}& = &{\tilde{\overline\Psi}}_{L_0}\gamma^\mu{\cal T}^a
\frac{(1-\sigma^3)}{2}{\tilde\Psi}_{L_0} +
\sum_{r= 2,4,...}{\tilde{\overline \Psi}}_{L_r}\gamma^\mu{\cal T}^a
{\tilde\Psi}_{L_r}\cr
\nonumber\\
&+&\sum_{s=1,3,...}{\tilde{\overline \Phi}}_{L_s}\gamma^\mu{\cal T}^a
{\tilde\Phi}_{L_s}.
\end{eqnarray}
As with conventional 
Pauli-Villars-Gupta regularization, the regularized composite current operator 
is summarized by
\begin{eqnarray}
\langle J^{\mu a}(x) \rangle_{reg}& = 
&\lim_{x \rightarrow y}Tr\{
-\gamma^\mu(x){\cal T}^a[
\frac{1}{2}(1-\sigma^3)
\langle T\{{\tilde\Psi}_{L_0}(x){\tilde{\overline\Psi}}_{L_0}(y)\}\rangle\cr 
\nonumber\\
&+&\sum_{r=2,4,...}
\langle T\{{\tilde\Psi}_{L_r}(x){\tilde{\overline\Psi}}_{L_r}(y)\}\rangle
+\sigma^3\sum_{s=1,3,...}
\langle T\{{\tilde\Phi}_{L_s}(x){\tilde{\overline\Phi}}_{L_s}(y)\} \rangle
]\}
\nonumber\\
\end{eqnarray}
with the trace running over Dirac and Yang-Mills indices. As a result
\begin{eqnarray}
\langle J^{\mu a}(x) \rangle_{reg}&= 
&\lim_{x \rightarrow y}Tr\{
\gamma^\mu(x){\cal T}^aP_L[
\frac{1}{2}(1-\sigma^3)
(i{{\cal D}\kern-0.15em\raise0.17ex\llap{/}\kern0.15em\relax})^{\dagger}
\frac{1}{(i{{\cal D}\kern-0.15em\raise0.17ex\llap{/}\kern0.15em\relax})
(i{{\cal D}\kern-0.15em\raise0.17ex\llap{/}\kern0.15em\relax})^{\dagger}} +\cr 
\nonumber\\
&+&\sum_{r=2,4,...}
(i{{\cal D}\kern-0.15em\raise0.17ex\llap{/}\kern0.15em\relax})^{\dagger}
\frac{1}{r^2\Lambda^2 + 
(i{{\cal D}\kern-0.15em\raise0.17ex\llap{/}\kern0.15em\relax})
(i{{\cal D}\kern-0.15em\raise0.17ex\llap{/}\kern0.15em\relax})^{\dagger}}
-\sum_{s=1,3,...}
(i{{\cal D}\kern-0.15em\raise0.17ex\llap{/}\kern0.15em\relax})^{\dagger}
\frac{1}{s^2\Lambda^2 + 
(i{{\cal D}\kern-0.15em\raise0.17ex\llap{/}\kern0.15em\relax})
(i{{{\cal D}\kern-0.15em\raise0.17ex\llap{/}\kern0.15em\relax}})^\dagger}]
\delta(x-y)\} \cr
\nonumber\\
&=&\lim_{x \rightarrow y}Tr\left\{\gamma^\mu(x)
{\cal T}^a\frac{1}{2}P_L\left[
\frac{1}{i{{\cal D}\kern-0.15em\raise0.17ex\llap{/}\kern0.15em\relax}}
\left(\sum_{n=-\infty}^{\infty}
\frac{(-1)^n{{\cal D}\kern-0.15em\raise0.17ex\llap{/}\kern0.15em\relax}
{{\cal D}\kern-0.15em\raise0.17ex\llap{/}\kern0.15em\relax}^\dagger}
{n^2\Lambda^2 + {{\cal D}\kern-0.15em\raise0.17ex\llap{/}\kern0.15em\relax}
{{\cal D}\kern-0.15em\raise0.17ex\llap{/}\kern0.15em\relax}^{\dagger}}
-\sigma^3\right)\right]\delta(x-y)\right\}\cr
\nonumber\\
&\equiv& \lim_{x \rightarrow y}Tr\left\{\gamma^\mu(x)
{\cal T}^a\frac{1}{2}P_L\left[
\frac{1}{i{{\cal D}\kern-0.15em\raise0.17ex\llap{/}\kern0.15em\relax}}
\left(f({{\cal D}\kern-0.15em\raise0.17ex\llap{/}\kern0.15em\relax}
{{\cal D}\kern-0.15em\raise0.17ex\llap{/}\kern0.15em\relax}^\dagger/\Lambda^2)
-\sigma^3\right)\right]\delta(x-y)\right\}.\cr 
\nonumber\\
\end{eqnarray}
In the above $n$ is summed over all integers, $i\Fslash \equiv  e^{1\over2}
i\Dslash e^{-{1\over2}}$ and $P_L \equiv \frac{1}{2}(1-\gamma^5)$.

The effect of the tower of regulators is to replace the divergent bare
 expression  
\begin{equation}
\langle J^{\mu a} \rangle_{bare} =
\lim_{x \rightarrow y}Tr\left\{\gamma^\mu(x)
{\cal T}^aP_L\left[
\frac{1}{i{{\cal D}\kern-0.15em\raise0.17ex\llap{/}\kern0.15em\relax}}
\frac{1}{2}\left(1-\sigma^3\right)\right]
\delta(x-y)\right\},
\end{equation}
by 
\begin{equation}
\langle J^{\mu a} \rangle_{reg}=
 \lim_{x \rightarrow y}Tr\left\{\gamma^\mu(x)
{\cal T}^a\frac{1}{2}P_L\left[
\frac{1}{i{{\cal D}\kern-0.15em\raise0.17ex\llap{/}\kern0.15em\relax}}
\left(f({{\cal D}\kern-0.15em\raise0.17ex\llap{/}\kern0.15em\relax}
{{\cal D}\kern-0.15em\raise0.17ex\llap{/}\kern0.15em\relax}^\dagger/\Lambda^2)
-\sigma^3\right)\right]\delta(x-y)\right\}.
\end{equation}
This general feature of the effect of the tower shows up in all the 
regularized currents.

The regulator function
\begin{eqnarray}
f(z) & \equiv & 
\sum_{n=-\infty}^{\infty}\frac{(-1)^n z}{n^2 + z} \cr
\nonumber\\
&=& \frac{\pi\sqrt{z}}{\sinh(\pi\sqrt{z})}
\end{eqnarray}
has the required properties \cite{FS, Okuyama} to ensure convergence. 
For instance, it falls rapidly to zero as $z \rightarrow \infty$. When 
the regulator masses are taken to $\infty, f(0)=1$. The
$\sigma^3$ part of the current remains unmodified, essentially 
 because the tower consists of regulators which are doubled in internal 
space and is  ``$\sigma^3$ neutral''.  It, therefore, can regularize 
only the singlet part of the $\frac{1}{2}(1-\sigma^3)$ projection of the 
bare current. However, when
the representation satisfies the anomaly cancellation conditions for
perturbative chiral gauge \cite{gglash} and mixed Lorentz-gauge anomalies
 \cite{nieh} i.e.
\begin{equation}
Tr\left( T^a\left\{T^b, T^c\right\}\right) =0
\end{equation}
and
\begin{equation}
Tr(T^a) =0
\end{equation}
respectively, 
the $\sigma^3$ part gives rise to no further divergences and, 
the current is successfully regularized by the tower 
of regulators. That all this is true is explained in Ref. \cite{inv}.

We turn next to the gravitational currents.
The bare spin current coupled to $A^-_{\mu AB}$ is
\begin{equation}
J^{\mu AB}=
{\tilde{\overline \Psi}}_{L_0}\gamma^\mu\frac{i}{2}\sigma^{AB}P_L
\frac{1}{2}(1-\sigma^3){\tilde\Psi}_{L_0}.
\end{equation}
Similarly, the regularized expression is
\begin{equation}
\langle J^{\mu AB} \rangle_{reg}=
 \lim_{x \rightarrow y}Tr\left\{\gamma^\mu(x)
{\frac{i}{2}\sigma^{AB}}\frac{1}{2}P_L\left[
\frac{1}{i{{\cal D}\kern-0.15em\raise0.17ex\llap{/}\kern0.15em\relax}}
\left(f({{\cal D}\kern-0.15em\raise0.17ex\llap{/}\kern0.15em\relax}
{{\cal D}\kern-0.15em\raise0.17ex\llap{/}\kern0.15em\relax}^\dagger/\Lambda^2)
-\sigma^3\right)\right]\delta(x-y)\right\}.
\end{equation}

Various proposals for 
defining the  energy momentum tensor have been suggested \cite{Fujienergy}.
If the classical bare action is regarded as
$S^-\left({\overline\Psi_{L}}, \Psi_{L}, e_{\mu A}, W_{\mu a}, A^-_{\mu AB}
\right)$ in the first line of Eq.(1),
then the energy momentum tensor $\Theta_{\mu\nu}$ is obtained 
from  
\begin{eqnarray}
e\Theta_{\mu\nu}&=& e_{\mu A}\frac{\delta S^-}{\delta E^\nu_A}\cr
\nonumber\\
&=&{\overline\Psi_{L}}{\gamma_\mu}iD_\nu\Psi_{L} -g_{\mu\nu}{\cal L},
\end{eqnarray}
where ${\cal L}$ is the Lagrangian.
On the other hand, if the variables 
${\tilde{\overline\Psi}_L}$ and ${\tilde\Psi}_L$ are to be treated as 
independent integration variables as is suggested by the
 diffeomorphism-invariant measure (20), 
then the energy momentum tensor $T_{\mu\nu}$
regarded as the source current for the vierbein is 
\begin{equation}
eT_{\mu\nu}= e_{\mu A}\frac{\delta{S}^-}{\delta E^\nu_A}
\end{equation}
with 
\begin{eqnarray}
{S}^-\left[{\tilde{\overline\Psi}_{L}}, {\tilde\Psi}_{L}, 
e_{\mu A}, W_{\mu a}, A^-_{\mu AB}\right] &=&
\int d^4x {\tilde{\overline\Psi}}_{L}e^{1\over2}i
{{D\kern-0.15em\raise0.17ex\llap{/}\kern0.15em\relax}}e^{-{1\over2}}
{\tilde\Psi}_{L} \cr
\nonumber\\
&=&\int d^4x {\tilde{\overline\Psi}}_{L}E^\mu_A\gamma^A
\left[iD_\mu -\frac{i}{2}(\partial_\mu \ln e)\right]\tilde\Psi_{L}.
\nonumber\\
\end{eqnarray}
The expression for the corresponding energy-momentum tensor is then
\footnote{In terms of variables which are not densitized,
$T_{\mu\nu}={{\overline\Psi}_{L}}{\gamma_\mu}iD_\nu{\Psi}_{L}
-\frac{i}{2}g_{\mu\nu}\left[\partial_\alpha(
{{\overline\Psi}}_{L}\gamma^\alpha{\Psi}_{L})+
\Gamma^\beta_{\beta\alpha}{{\overline\Psi}}_{L}\gamma^\alpha{\Psi}_{L})
\right].$}
\begin{equation}
eT_{\mu\nu}=
{\tilde{\overline\Psi}_{L}}{\gamma_\mu}
i\left(D_\nu -\frac{1}{2}\Gamma^\alpha_{\alpha\nu}\right){\tilde\Psi}_{L}
-\frac{i}{2}g_{\mu\nu}
\partial_\alpha({\tilde{\overline\Psi}}_{L}\gamma^\alpha{\tilde\Psi_{L}}).
\nonumber\\
\end{equation}
As a result, $T_{\mu\nu}$ and $\Theta_{\mu\nu}$ are related by
\begin{equation}
eT_{\mu\nu} = e\Theta_{\mu\nu} + \frac{1}{2}g_{\mu\nu}
\left({\tilde{\overline\Psi}}_{L}\frac{\delta{S}^-}
{\delta{\tilde{\overline\Psi}}_{L}} + {\tilde\Psi}_{L}
\frac{\delta{S}^-}{\delta{\tilde\Psi}_{L_0}}\right).
\end{equation}
The difference between the two is therefore not significant classically 
when the equations of motion can be imposed. However, at the quantum level, 
there can be subtleties \cite{Fujienergy}. 
Due to the choice of the densitized variables, all bare mass terms and, in 
particular, regulator bare mass terms, are independent of the vierbein and 
therefore do {\it not} contribute to $T_{\mu\nu}$. 
The regularized expression consequently becomes
\begin{eqnarray}
\langle eT_{\mu\nu}\rangle_{reg}&=&
\lim_{x \rightarrow y}Tr\left\{{\gamma_\mu}
i\left(D_\nu -\frac{1}{2}\Gamma^\alpha_{\alpha\nu}\right)
\frac{1}{i{{\cal D}\kern-0.15em\raise0.17ex\llap{/}\kern0.15em\relax}}P_L 
\frac{1}{2}
\left[f\left(\frac{{{\cal D}\kern-0.15em\raise0.17ex\llap{/}\kern0.15em\relax}
{{\cal D}\kern-0.15em\raise0.17ex\llap{/}\kern0.15em\relax}^\dagger}
{\Lambda^2}\right)
-{\sigma^3}\right]\delta(x-y)
\right\} \cr
\nonumber\\
&-&\frac{i}{2}g_{\mu\nu}\langle\partial_\alpha J^\alpha \rangle_{reg},
\end{eqnarray}
where $-J^\alpha = J^\alpha_5$ is the ABJ current.

Again, the $\sigma^3$ part of 
the energy-momentum tensor gives rise to no divergent fermion loops if 
conditions (58) and (59) hold. 
Hence the 
expression for the energy-momentum tensor is regularized for finite $\Lambda$.

The regularized trace of the energy-momentum tensor is therefore
\begin{equation}
\langle eT^\mu\,_\mu\rangle_{reg}=
\lim_{x \rightarrow y}Tr\left\{P_L \frac{1}{2}
\left[f\left(\frac{{{\cal D}\kern-0.15em\raise0.17ex\llap{/}\kern0.15em\relax}
{{\cal D}\kern-0.15em\raise0.17ex\llap{/}\kern0.15em\relax}^\dagger}
{\Lambda^2}\right)
-{\sigma^3}\right]\delta(x-y)
\right\}
- 2i\langle\partial_\mu J^\mu \rangle .
\end{equation}

In our present discussion, we do not densitize the 
{\it background} variables and eschew use, for instance, of
$W_{Aa}\equiv e^{1\over2}E^\mu_AW_{\mu a}$ instead of $W_{\mu a}$. This choice
would be useful if an explicitly diffeomorphism invariant measure 
$\prod DW_{Aa}$ is required when the path integral formalism is to be 
applied to the quantization of the gauge fields \cite{Fujienergy}. 

The energy-momentum tensor
should be symmetrized if it is to be regarded as 
the source of the metric.
Here in the first order formulation, the antisymmetric part of the
energy-momentum tensor serves as the source of the spin current (see also 
Eq.(84)).
It is known that there are no perturbative Lorentz anomalies in four 
dimensions \cite{Chang}. This is {\it verified by the explicitly 
Lorentz-invariant regularization scheme} proposed here and can 
indeed be checked
by the explicit verification of the consistency 
condition for the absence of Lorentz anomalies. 
This condition is derived later on in Section V.

To summarize, all the currents carry left-handed projections and are
regularized by the scheme for finite $\Lambda$. Fermion loops are
generated by the multipoint correlation functions obtained by
functional differentiation of the regularized currents in Eqs. (56), (61)
and (67). We may further note the explicit role of the ABJ anomaly in 
Eqs. (67) and (68).

Under a chiral $\gamma^5$ rotation, 
\begin{eqnarray}
{\tilde\Psi}_{L_r} \rightarrow 
e^{i\alpha\gamma^5}{\tilde\Psi}_{L_r} &=& e^{-i\alpha}{\tilde\Psi}_{L_r},\cr
\nonumber\\  
{\tilde{\overline\Psi}}_{L_r} \rightarrow 
{\tilde{\overline\Psi}}_{L_r} e^{i\alpha\gamma^5}
&=&{\tilde{\overline\Psi}}_{L_r}e^{i\alpha}, \label{eq:abj}
\end{eqnarray}
and similarly for ${\tilde\Phi}_{L_s}$ and ${\tilde{\overline\Phi}}_{L_s}$.
Kinetic terms are invariant under this global tranformation, but mass terms 
are not.
The bare massless action is invariant under such a global transformation, and
 the associated ABJ or $\gamma^5$ current 
\begin{equation} 
J^\mu_5 = 
{\tilde{\overline\Psi}}_{L_0}\gamma^\mu\gamma^5\tilde\Psi_{L_0}
= -{\tilde{\overline\Psi}}_{L_0}\gamma^\mu\tilde\Psi_{L_0}
=-J^\mu_F,
\end{equation}
is conserved classically, i.e.. $\partial_\mu J^\mu_5 = 0.$
However, the bare quantum composite current 
\begin{equation}
\langle J^{\mu}_5 \rangle_{bare} =
-\lim_{x \rightarrow y}Tr\left\{\gamma^\mu(x)P_L\left[
\frac{1}{i{{\cal D}\kern-0.15em\raise0.17ex\llap{/}\kern0.15em\relax}}\frac{1}
{2}\left(1-\sigma^3\right)\right]
\delta(x-y)\right\}
\end{equation}
is divergent. The regularized current is not necessarily conserved. 
In the generalized Pauli-Villars-Gupta scheme, the mass terms of the 
regulators break the symmetry explicitly. 
The expectation value of the regularized ABJ current is
\begin{equation}
\langle J^{\mu}_5(x) \rangle_{reg}
=-\lim_{x \rightarrow y}Tr\left\{\gamma^\mu(x)
\frac{1}{2}(1-\gamma^5)
{\frac{1}{i{{\cal D}\kern-0.15em\raise0.17ex\llap{/}\kern0.15em\relax}}}
\frac{1}{2}\left(f({{\cal D}\kern-0.15em\raise0.17ex\llap{/}\kern0.15em\relax}
{{\cal D}\kern-0.15em\raise0.17ex\llap{/}\kern0.15em\relax}^\dagger/\Lambda^2)
-\sigma^3\right)\delta(x-y)\right\}.  
\end{equation} 
The previous arguments concerning the unregulated $\sigma^3$ 
part are still valid.  Within this context, we have in effect 
regularized  the ABJ current,
and the associated amplitudes can be computed explicitly.

The ABJ anomaly 
can be explicitly computed by taking the divergence 
of the expectation value of the regularized expression in Eq.(72) as 
\begin{equation}
\langle \partial_\mu J^\mu_5 \rangle_{reg} =
\partial_\mu \lim_{x \rightarrow y}Tr\left\{-\gamma^\mu
\frac{1}{2}(1-\gamma^5)
{\frac{1}{i{{\cal D}\kern-0.15em\raise0.17ex\llap{/}\kern0.15em\relax}}}
\frac{1}{2}\left(f({{\cal D}\kern-0.15em\raise0.17ex\llap{/}\kern0.15em\relax}
{{\cal D}\kern-0.15em\raise0.17ex\llap{/}\kern0.15em\relax}^\dagger/\Lambda^2)
-\sigma^3\right)\delta(x-y)\right\}. 
\end{equation} 
The trace can be evaluated by using the complete sets of 
eigenvectors, $\{X_n\}$ and $\{Y_n\}$, of the 
positive-semidefinite Hermitian operators with
\begin{eqnarray}
{{\cal D}\kern-0.15em\raise0.17ex\llap{/}\kern0.15em\relax}
{{\cal D}\kern-0.15em\raise0.17ex\llap{/}\kern0.15em\relax}^\dagger X_n &=& 
\lambda^2_n X_n, \cr
\nonumber\\ 
{{\cal D}\kern-0.15em\raise0.17ex\llap{/}\kern0.15em\relax}^\dagger
{{\cal D}\kern-0.15em\raise0.17ex\llap{/}\kern0.15em\relax} Y_n &=& 
\lambda^2_n Y_n.
\end{eqnarray}
Consequently,  
\begin{eqnarray}
\langle \partial_\mu J^\mu_5 \rangle_{reg} 
&=&\lim_{\Lambda \rightarrow \infty}
\frac{i}{4}Tr[\gamma^5f
({{\cal D}\kern-0.15em\raise0.17ex\llap{/}\kern0.15em\relax}^\dagger
{{\cal D}\kern-0.15em\raise0.17ex\llap{/}\kern0.15em\relax}/\Lambda^2)
+ \gamma^5f
({{\cal D}\kern-0.15em\raise0.17ex\llap{/}\kern0.15em\relax}
{{\cal D}\kern-0.15em\raise0.17ex\llap{/}\kern0.15em\relax}^\dagger/
\Lambda^2)]\cr
\nonumber\\
&=&\lim_{\Lambda \rightarrow \infty}
\frac{i}{4}\sum_n[Y^\dagger_n\gamma^5f
({{\cal D}\kern-0.15em\raise0.17ex\llap{/}\kern0.15em\relax}^\dagger
{{\cal D}\kern-0.15em\raise0.17ex\llap{/}\kern0.15em\relax}/\Lambda^2)Y_n
+ X^\dagger_n\gamma^5f
(`{{\cal D}\kern-0.15em\raise0.17ex\llap{/}\kern0.15em\relax}
{{\cal D}\kern-0.15em\raise0.17ex\llap{/}\kern0.15em\relax}^\dagger/
\Lambda^2)X_n].
\nonumber\\
\end{eqnarray}

For Euclidean signature, this works out to be
\begin{equation}
\langle \partial_\mu J^\mu_5 \rangle
={{i\times d}\over{768\pi^2}}F_{\alpha\beta AB}
\epsilon^{\alpha\beta\mu\nu}F_{\mu\nu}\,^{AB}
+{i\over{32\pi^2}}Tr(\epsilon^{\alpha\beta\mu\nu}
G_{\alpha\beta a}T^aG_{\mu\nu b}T^b)
\end{equation}
in the absence of torsion.\footnote{When there is torsion, an 
additional contribution which diverges as the regulator masses are taken to 
infinity is present.
 The associated counterterm  will be discussed in Section VII.}
 In the above,
$G_{\mu\nu a}$ and $F_{\mu\nu AB}$ are, respectively, the curvatures of
$W_{\mu a}$ and $A_{\mu AB}$.
This gives the result which is {\it one-half} of the chiral anomaly of a 
vector theory. Because all the fields are Weyl, the factor we get for
the gravitational part is also $d$ rather than 2$d$. This is in agreement with
the fact that there are $d$ Weyl fermions coupled to gravity in the bare 
action.

\bigskip
%newstuff
\section*{V. EFFECTIVE ACTION AND CONSERVATION EQUATIONS}
\bigskip

In order to derive the conservation equations associated with the local
symmetries of the quantum theory, we may consider the generating function 
\begin{eqnarray}
Z &=& \exp(-\Gamma_{eff.}[E^{\mu A}, A^-_{\mu AB}, W_{\mu a}])\cr
\nonumber\\
&=& \int D{\tilde{\overline\Psi}}_L D{\tilde\Psi}_L
\exp(-{S}^-[{\tilde{\overline\Psi}}_L, {\tilde\Psi}_L, E^{\mu A},
A^-_{\mu AB}, W_{\mu a}]).
\nonumber\\
\end{eqnarray}
Under a change of integration variables,
\begin{equation}
{\tilde\Psi}_L \rightarrow {\tilde\Psi}'_L, \qquad
{\tilde{\overline\Psi}}_L \rightarrow {\tilde{\overline\Psi}}'_L ,
\end{equation}
there is no change in the partition function if there are {\it no anomalous
Jacobians} in the measure.\footnote{The measure may be defined by expansion
in terms of the complete sets of eigenvectors $\left\{X_n \right\}$ and 
$\left\{Y_n \right\}$ as in Eq.(74).} 
Thus
\begin{equation}
0=\delta\ln Z = 
-\int_M (\langle 
\delta{\tilde{\overline\Psi}}_L
{{\delta{S}^-}\over{\delta{\tilde{\overline\Psi}}_L}} 
+ \delta{\tilde{\Psi}}_L
{{\delta{S}^-}\over{\delta{\tilde{\Psi}}_L}}\rangle).
\end{equation}
But under simultaneous transformations
\begin{eqnarray} 
E^{\mu A} &\rightarrow& E'^{\mu A}\cr
\nonumber\\
A_{\mu AB} &\rightarrow& A'_{\mu AB}\cr
\nonumber\\
W_{\mu a} &\rightarrow& W'_{\mu a}\cr
\nonumber\\
{\tilde\Psi}_L &\rightarrow& {\tilde\Psi}'_L, \qquad
{\tilde{\overline\Psi}}_L \rightarrow {\tilde{\overline\Psi}}'_L 
\end{eqnarray}
which correspond to {\it symmetries} of the action,
\begin{equation}
\delta S^-=
\int_M (\delta{\tilde{\overline\Psi}}_L
{{\delta{S}^-}\over{\delta{\tilde{\overline\Psi}}_L}} 
+ \delta{\tilde{\Psi}}_L{{\delta{S}^-}\over{\delta{\tilde{\Psi}}_L}}
+\delta E^{\mu A}{{\delta{S}^-}\over{\delta E^{\mu A}}} 
+ \delta A_{\mu AB}
{{\delta{S}^-}\over{\delta A_{\mu AB}}}
+\delta W_{\mu a}{{\delta{S}^-}\over{\delta W_{\mu a}}}) =0.
\end{equation}
Therefore for such symmetry transformations
\begin{eqnarray}
-\delta \Gamma_{eff.}=\delta \ln Z &=&\int_M(\langle 
\delta E^{\mu A}{{\delta{S}^-}\over{\delta E^{\mu A}}} 
+ \delta A_{\mu AB}
{{\delta{S}^-}\over{\delta A_{\mu AB}}}
+\delta W_{\mu a}{{\delta{S}^-}\over{\delta W_{\mu a}}} \rangle)\cr
\nonumber\\
&=& -\int_M (
{{\delta\Gamma_{eff.}}\over{\delta E^{\mu A}}}\delta E^{\mu A}
+ {{\delta\Gamma_{eff.}}\over{\delta A_{\mu AB}}}\delta A_{\mu AB}
+{{\delta\Gamma_{eff.}}\over{\delta W_{\mu a}}}\delta W_{\mu a})\cr
\nonumber\\
&=&-\int_M (
\langle E^{\nu A}eT_{\nu\mu}\rangle \delta E^{\mu}_A +
\langle J^{\mu AB}\rangle \delta A_{\mu AB}+
\langle J^{\mu a}\rangle \delta W_{\mu a})\cr
\nonumber\\
&=&0,
\nonumber\\
\end{eqnarray}
since
\begin{eqnarray}
\langle E^{\mu A}eT_{\mu\nu}\rangle &=&
{{\delta\Gamma_{eff.}}\over{\delta E^{\nu A}}},\cr
\nonumber\\
\langle J^{\mu AB}\rangle &=&{{\delta\Gamma_{eff.}}
\over{\delta A_{\mu AB}}},\cr
\nonumber\\
\langle J^{\mu a}\rangle &=&
{{\delta\Gamma_{eff.}}\over{\delta W_{\mu a}}}.
\nonumber\\
\end{eqnarray}

The resultant conservation equations for local gauge, Lorentz 
and diffeomorphism symmetries are, respectively,
\begin{eqnarray}
\langle D_\mu J^{\mu a}\rangle &=&0,\cr
\nonumber\\
E^{[\mu A}E^{\nu B]}\langle eT_{\mu\nu}\rangle +
\langle D_\mu J^{\mu AB}\rangle &=&0 
\nonumber\\
\end{eqnarray}
and
\begin{eqnarray}
&&-A^-_{\alpha{AB}}\left\{ 
E^{[\mu A}E^{\nu B]}\langle eT_{\mu\nu} \rangle +
\langle D_\mu J^{\mu AB}\rangle \right\}
-\langle W_{\alpha a} D_\mu J^{\mu a}\rangle \cr
\nonumber\\
&& +\langle 
\partial_\mu(eT^\mu\,_\alpha) - e\Gamma^\nu_{\alpha\mu}T^\mu\,_\nu
\rangle
+ \langle G_{\alpha \mu a} J^{\mu a}\rangle
+ \langle F^-_{\alpha\mu AB} J^{\mu AB}\rangle =0
\nonumber\\
\end{eqnarray}
with  $F^-_{\mu\nu AB}$ and $G_{\mu\nu a}$ being the respective curvatures of
$A^-_{\mu AB}$ and $W_{\mu a}$.

The first equation is just the condition for the gauge current to be
conserved. In the second equation, note that the antisymmetric part of 
the energy-momentum tensor acts as the source of the spin current. 
The final equation is the complete expression for invariance under local
infinitesimal diffeomorphsims in the first order formulation when there are 
also couplings to internal gauge fields. This expression is more involved
the familiar condition
\begin{equation}
\langle \nabla_\mu T^\mu\,_\alpha\rangle =0
\end{equation}
for scalar fields when the gravitational coupling is only through the metric.
The expression given here for fermion theories in the first order 
formulation agrees with that of 
Nieh and Yan \cite{nay}. Note that
\begin{equation}
\partial_\mu(eT^\mu\,_\alpha) - e\Gamma^\nu_{\alpha\mu}T^\mu\,_\nu
= e(\nabla_\mu T^\mu\,_\alpha + 2B_\mu T^\mu\,_\alpha).
\end{equation}
We may also observe that all the currents defined here 
are left-handed and therefore 
only the left-handed spin connection is projected in Eqs. (84) and (85).

Equipped with a regularization scheme, 
we can check explicitly that these equations for
the expectation values which ensure the local symmetries of the
theory are free of anomalies, are indeed satisfied. In particular, we can
check that the equations for Lorentz and diifeomorphism invariance 
hold thus ensuring {\it no inconsistencies or 
perturbative anomalies for the non-Hermitian Weyl theory}
studied here, {\it despite} the unfamiliar appearances of the ABJ current
and ABJ anomaly in the imaginary parts of the spin current and 
energy-momentum tensor.
This is in agreement with the fact that an explicitly gauge as well as
Lorentz and diffeomorphism invariant regularization scheme can be found
for the theory \cite{inv}.

In the next section we shall consider the case of global ABJ and scaling
symmetries when there are anomalous Jacobians in the measure. Note that
the Pauli-Villars regularization scheme discussed earlier explicitly
preserves the local symmetries of the theory but also explicitly violate
 global scaling and $\gamma^5$ transformations
since the regulator mass terms are not invariant under these. 
Thus the scheme also provides consistent computations of these anomalies for 
the Weyl theory.

\bigskip
\section*{VI. ANOMALOUS SYMMETRIES}
\bigskip

In addition to the local gauge, Lorentz and diffeomorphism symmetries, the
bare action is also invariant under global $\gamma^5$ and scaling 
transformations. Classically this results in the conservation of the
ABJ current and the tracelessness of the energy-momentum tensor.
However, quantum mechanically these symmetries are violated by the 
regularization which breaks these symmetries explicitly. 

In the path integral approach, the anomalies are related to the nontrivial
Jacobians of the fermion measure \cite{fuji}.  
To obtain the relation between the
anomalous Jacobians and the expectation values of the currents, we recall
that the partition function $Z$ is invariant under an arbitrary change of 
fermion integration variables. However if there is an anomalous
Jacobian in the measure which transforms as
\begin{equation}
D{\tilde{\overline\Psi}}_L D{\tilde\Psi}_L
\rightarrow 
D{\tilde{\overline\Psi}}'_L D{\tilde\Psi}'_L=
\exp({\cal A})D{\tilde{\overline\Psi}}_L D{\tilde\Psi}_L,
\end{equation}
then 
\begin{equation}
Z= \int \exp({\cal A})D{\tilde{\overline\Psi}}_L D{\tilde\Psi}_L
\exp(-S^-[{\tilde{\overline\Psi}}'_L, {\tilde\Psi}'_L, e_A, A^-_{AB}, W_a]),
\end{equation}
since the partition function is unaltered by a change of 
integration variables.

For infinitesimal transformations, we have
\begin{eqnarray}
0&=&\delta Z\cr
\nonumber\\
&=& ({\cal A})Z -  
\int D{\tilde{\overline\Psi}}_L D{\tilde\Psi}_L
\int_M (\delta{\tilde{\overline\Psi}}_L
{{\delta{S}^-}\over{\delta{\tilde{\overline\Psi}}_L}} 
+ \delta{\tilde{\Psi}}_L{{\delta{S}^-}\over{\delta{\tilde{\Psi}}_L}})
\exp(-S^-[{\tilde{\overline\Psi}}_L, {\tilde\Psi}_L, e_A, A^-_{AB}, W_a]).
\nonumber\\
\end{eqnarray}
This yields the sought-after relation
\begin{equation}
{\cal A}  = 
\langle
\int_M (\delta{\tilde{\overline\Psi}}_L
{{\delta{S}^-}\over{\delta{\tilde{\overline\Psi}}_L}} 
+ \delta{\tilde{\Psi}}_L{{\delta{S}^-}\over{\delta{\tilde{\Psi}}_L}})
\rangle.
\end{equation}

In the case of $\gamma^5$ transformations, 
\begin{equation}
\delta{\tilde{\overline\Psi}}_L = i\alpha{\tilde{\overline\Psi}}_L,
\qquad 
\delta{\tilde{\Psi}}_L = -i\alpha{\tilde{\Psi}}_L,
\end{equation}
and
\begin{equation}
-i\int_M \alpha A^5 = 
\langle
\int_M i\alpha({\tilde{\overline\Psi}}_L
{{\delta{S}^-}\over{\delta{\tilde{\overline\Psi}}_L}} 
- {\tilde{\Psi}}_L{{\delta{S}^-}\over{\delta{\tilde{\Psi}}_L}})
\rangle
\end{equation}
with ${\cal A}$ written as $-i\int_M \alpha A^5$.
The ABJ anomaly due to the nontrivial Jacobian is therefore
\begin{equation}
A^5= -\langle i\partial_\mu ({\tilde{\overline\Psi}}_L\gamma^\mu{\tilde\Psi}_L)
\rangle. 
\end{equation}
Given a regularization scheme, we can actually evaluate the anomaly as
the expectation value of the divergence  of the regularized current.

In the case of scaling,
\begin{equation}
\delta{\tilde{\overline\Psi}}_L = \frac{1}{2}\alpha{\tilde{\overline\Psi}}_L,
\qquad 
\delta{\tilde{\Psi}}_L = \frac{1}{2}\alpha{\tilde{\Psi}}_L,
\end{equation}
and writing ${\cal A}= \int_M \alpha A$ yields
\begin{equation}
\int_M \alpha A = {1\over2}\langle
\int_M \alpha({\tilde{\overline\Psi}}_L
{{\delta{S}^-}\over{\delta{\tilde{\overline\Psi}}_L}} 
+ {\tilde{\Psi}}_L{{\delta{S}^-}\over{\delta{\tilde{\Psi}}_L}})
\rangle.
\end{equation}
Hence the anomaly associated with the nontrivial Jacobian under scaling is
\begin{equation}
A= {1\over2}\langle\left(
{\tilde{\overline\Psi}}_Le^{1\over2}i\Dslash e^{-{1\over2}} 
{\tilde\Psi}_L -i\partial_\mu({\tilde{\overline\Psi}}_L
\gamma^\mu){\tilde\Psi}_L +{\tilde{\overline\Psi}}_L
\gamma^\mu({i\over2}A_{\mu AB}\sigma^{AB}-\frac{i}{2}\partial_\mu \ln e +
 W_{\mu a}T^a){\tilde\Psi}_L
\right)\rangle.
\end{equation}
By comparing with the trace of Eq. (65), we see that the anomaly from
the nontrivial Jacobian under scaling of the integration variables is related 
to the trace of the energy-momentum tensor (see also Eq.(107)) by
\begin{equation}
A= \langle eT^\mu\,_\mu + i\frac{3}{2}\partial_\mu J^\mu\rangle.
\end{equation}
There is an additional term proprotional to the divergence of the ABJ
current which is nonvanishing quantum mechanically.
This relation can also be deduced by considering the simultaneous local
scaling transformations as in (26) and (27). Under such a scaling the effective
action $\Gamma_{eff.}[e_A, A^-_{AB}, W_{a}]$ changes by
\begin{eqnarray}
\delta \Gamma_{eff.}& =& 
\int_M \langle eT_{\nu\mu} E^{\nu A}\rangle
\delta E^\mu\,_A \cr
\nonumber\\
&=&\int_M \langle eT_{\nu\mu} E^{\nu A}\rangle
g^{\nu\alpha}e_{\alpha}\,^A\delta E^\mu\,_ A \cr
\nonumber\\
&=&-\frac{1}{4}\int_M \langle eT^\mu\,_\mu \rangle\delta (\ln e) \cr
\nonumber\\
&=&-\int_M \langle \alpha eT^\mu\,_\mu \rangle
\nonumber\\
\end{eqnarray}
since $\delta (\ln e) = 4\alpha$ under scaling. But the Weyl action changes by
\begin{equation}
S^- \rightarrow S^- -i\frac{3}{2}\int_M (\partial_\mu \alpha)J^\mu
\end{equation}
under local scale transformations. Thus, under simultaneous scale 
transformations we have
\begin{equation}
\exp(-\Gamma_{eff.}[\exp({\alpha})e_A, A^-_{AB}, W_a])
=\int \exp(\int_M \alpha A)D{\tilde{\overline\Psi}}_LD{\tilde\Psi}_L
\exp(-S^- -i\frac{3}{2}\int_M\alpha\partial_\mu J^\mu).
\end{equation}
For infinitesimal transformations, the result is therefore
\begin{eqnarray}
\langle eT^\mu\,_\mu\rangle& =&-{{\delta \Gamma_{eff.}} 
\over {\delta \alpha}}\cr
\nonumber\\
&=& A  + i\frac{3}{2} A^5
\nonumber\\
\end{eqnarray}
which is in complete agreement with the relation derived earlier.
$A$ can be computed from the regularized expressions of
$\langle eT^\mu\, _\mu \rangle$ and $\langle \partial_\mu J^\mu \rangle$.

Had the Hermitized Weyl action been used, then  under local scaling
the action is invariant, and the result will be that
the anomalous Jacobian agrees with the trace of the energy-momentum with 
no extra imaginary ABJ-anomaly contribution. 
Another way to understand this is to note that 
as far as scaling is concerned, the Weyl action
differs from the Hermitized version by a scale-noninvariant term 
$-i\int_M B_\mu J^\mu$ which transforms as
\begin{equation}
-i\int_M B_\mu J^\mu \rightarrow -i\int_M B_\mu J^\mu -i\frac{3}{2}\int_M
(\partial_\mu \alpha)J^\mu
\end{equation}
since $B_\mu$ changes by $\frac{3}{2}\partial_\mu \alpha$ while $J^\mu$ is 
invariant under local scaling. Thus these differences in scaling behaviour 
from the Hermitized theory offer further physical avenues to test the validity 
of the Weyl theory.

%endnewstuff
\newpage
\bigskip
\section*{VII. CPT VIOLATION}
\bigskip

In order to couple fermions to the four forces in a completely chiral
fashion without introducing the right-handed spin connection
we do not Hermitize the Weyl action. The difference 
between the Weyl action and the Hermitian version,
as has been addressed in Sections II and III, involves the divergence
of the ABJ current and also torsion terms. This difference is {\it subtle} 
because classically, the ABJ current is conserved while torsion is also
zero for most familiar background solutions in general relativity. 
Moreover, within the context of Section III, $B_\mu$ is zero on-shell.
However, quantum mechanically, the ABJ anomaly exists and off-shell torsion 
in the presence of fermions cannot be guaranteed to vanish.
As a result, among other things, the energy-momentum tensor and spin current
presented here acquire imaginary terms (in Lorentzian signature 
spacetimes). These originate precisely
from the non-Hermiticity of the Weyl Lagrangian 
whose anti-Hermitian part is not of the form
of a ordinary divergence but has local contributions when there is torsion.
Since the expectation value of the divergence of the ABJ current is not 
zero quantum mechanically, there are subtle violations of
discrete symmetries due to the ABJ current and ABJ anomaly in the 
presence of topologically nontrivial gauge and gravitational instantons, 
and also nonvanishing torsion.  It is easy to check that the
{\it imaginary} part of the Weyl action is CP and {\it CPT-odd} since it is 
local and Lorentz invariant.\footnote{Details can also be found in Refs.
\cite{soo, cps}.} Thus the Weyl action while obeying all the
local gauge, Lorentz and diffeomorphism symmetries of the theory
violates discrete symmetries such as CP and CPT in contrast with
the Hermitized Weyl and Majorana theories.
 
In addition to possible nonperturbative violations due to instantons,
the precise {\it perturbative and local} processes
which are involved are contained in the imaginary parts of the spin
current and the energy-momentum tensor and the fermion loops 
generated by them. All these processes involve torsion and the ABJ current
and originate from the $B_\mu J^\mu$ coupling in the imaginary
part of the action.

To isolate these processes we decompose the energy-momentum and 
spin current into their Hermitian and anti-Hermtian parts.
\begin{eqnarray}
eT_{\mu\nu}&=& {\tilde{\overline\Psi}}_L\gamma_\mu[iD_\mu 
-\frac{i}{2}(\partial_\mu\ln e)]{\tilde\Psi}_L 
-\frac{i}{2}g_{\mu\nu}\partial_\alpha J^\alpha \cr
\nonumber\\
&=&\frac{1}{2}(eT_{\mu\nu}+ (eT_{\mu\nu})^\dagger) + iIm(eT_{\mu\nu})
\nonumber\\
\end{eqnarray}
where
\begin{eqnarray}
\frac{1}{2}(eT_{\mu\nu}+ (eT_{\mu\nu})^\dagger) & =&
{\tilde{\overline\Psi}}_L\gamma_\mu[iD_\nu 
-\frac{i}{2}(\partial_\nu\ln e)]{\tilde\Psi}_L  \cr
\nonumber\\
&&-i(\partial_\nu{\tilde{\overline\Psi}}_L)
\gamma_\mu{\tilde\Psi}_L +\frac{i}{2}{\tilde{\overline\Psi}}_L[
(\partial_\nu\ln e) +A_{\nu AB}\sigma^{AB}]\gamma_\mu{\tilde\Psi}_L
\nonumber\\
\end{eqnarray}
and 
\begin{equation}
iIm(eT_{\mu\nu}) = -\frac{i}{2}g_{\mu\nu}\partial_\alpha J^\alpha
+\frac{i}{2}\left[\partial_\nu J_\mu
 -\Gamma^\alpha_{\mu\nu}J_\alpha -\Gamma^\alpha_{\alpha\nu}J_\mu
\right]
\end{equation}
with $J_\mu = g_{\mu\nu}{\tilde{\overline\Psi}}_L\gamma^\nu{\tilde\Psi}_L$.

This implies that the trace of the energy-momentum tensor which is related to 
the conformal anomaly also picks up an imaginary term. The full expression
is
\begin{eqnarray}
eT^\mu\,_\mu&=&{\tilde{\overline\Psi}}_L\gamma^\mu\left[
i\partial_\mu + \frac{i}{2}A_{\mu AB}\sigma^{AB}-\frac{i}{2}\partial_\mu\ln e
+W_{\mu a}T^a \right]{\tilde\Psi}_L 
-2i\partial_\mu({\tilde{\overline\Psi}}_L\gamma^\mu{\tilde\Psi}_L )\cr 
\nonumber\\
&=&\frac{1}{2}({\tilde{\overline\Psi}}_L\gamma^\mu e^{1\over2}
iD_\mu e^{-{1\over2}}{\tilde\Psi}_L + H. c.)
-i\frac{3}{2}\partial_\mu J^\mu - iB_\mu J^\mu.
\nonumber\\
\end{eqnarray}

The spin current
\begin{equation}
J^{\mu AB} = \frac{i}{2}{\tilde{\overline\Psi}}_L
\gamma^\mu\sigma^{AB}{\tilde\Psi}_L
\end{equation}
has an anti-Hermitian part which is
\begin{eqnarray}
iIm(J^{\mu AB})&=& \frac{i}{4}{\tilde{\overline\Psi}}_L\left[
\gamma^B E^{\mu A} - \gamma^A E^{\mu B}\right]{\tilde\Psi}_L \cr
\nonumber\\
&=& \frac{i}{4} \left[e_\nu\,^B E^{\mu A}- e_\nu\,^A E^{\mu B}\right]J^\nu
\nonumber\\
\end{eqnarray}

All these imaginary terms are {\it not present} in the conventional
Hermitian theory. We see that there are CPT-violating 
processes from fermion loops generated by the fermion 
(hence axial or ABJ) current coupled to $B_\mu$. 
As we have shown, this ABJ current, although anomalous,
is regularized by the proposed scheme. 
In principle, the discussion provided here gives a 
self-consistent and self-contained method to compute these processes
by computing the expectation values of the imaginary parts of the spin 
current and energy-momentum tensor in background fields.

We can estimate some of the effects of these {\it new} processes on the 
effective action. 
It can be argued that when the standard model {\it chirally} 
coupled to gravity is quantized in curved spacetimes 
with nonvanishing torsion, then the one-loop effective action will necessarily
yield the Samuel-Jacobson-Smolin action with cosmological constant as the 
lowest order curvature counterterm just as the Hermitian 
Weyl action for spin 1/2 particles must require an effective action which 
contains the cosmological and Einstein-Hilbert-Palatini actions as the 
lowest order counterterms if we include gravitational couplings of
 both chiralities. Thus a quantum field theory of Weyl fermions in background
curved spacetimes coupled in this chiral manner must violate CPT to the 
lowest order in curvature by a term of the form 
${i/({16\pi G_{renor.}})}\int_M e^A\wedge e^B\wedge F_{AB}$ in the 
effective action. 

We sketch the arguments
\footnote{Details of fermion loop calculations using the  
explicitly invariant chiral regularization will be presented elsewhere.} 
of how and why for the truly Weyl theory
such an imaginary term in addition to the familiar real counterterms 
appears in the effective action. As emphasized, 
the Weyl fermion action contains no coupling to the right-handed or
self-dual spin connection $A^+_{AB} = \frac{1}{2}(iA_{AB} 
+\frac{1}{2}\epsilon_{AB}\,^{CD} A_{CD})$. So the usual 
Einstein-Hilbert-Palatini action which involves both $A^+_{AB}$ and $A^-_{AB}$
cannot occur as a counterterm in the effective action without 
modification. This is particularly 
clear if an explicitly chiral regularization which also involves 
only $A^-_{AB}$ but no $A^+_{AB}$ spin connection such the one advocated 
here is used. Moreover, whenever the spin 
connection makes its appearance in a counterterm, it must appear only 
in the anti-self-dual combination of $A^-_{AB}$. The lowest order curvature
term invariant under all the local chiral symmetries of the theory is the 
Samuel-Jacobson-Smolin action\footnote{The cosmological constant term is real 
and appears as one of the usual counterterms generated by quantized fermions.}
which is the anti-self-dual projection of the Einstein-Hilbert-Palatini
action. 
To be more explicit, we can also consider the CPT-violating terms in the
fermionic Weyl action as in Eq. (24) and relate 
the counterterms to the imaginary pieces of the Samuel-Jacobson-Smolin action
which will be generated by the imaginary Weyl action in addition to the usual
counterterms. \footnote{It may be worth pointing out that with nonvanishing 
torsion, the complete list of counterterms for even the ordinary Hermitized 
theory is quite involved. See for instance Ref. \cite{nay}.}  
For this purpose, we may note that
\begin{equation}
{i\over{16\pi G}}\int_M e^A\wedge e^B\wedge F_{AB} = 
-{i\over{16\pi G}}\int_M \left\{ d(e^A\wedge T_A) - T_A\wedge T^A\right\}.
\end{equation}
The imaginary boundary term from the integral of ABJ-anomaly in Eq. (24) is 
related to the first boundary term above. $d(e^A\wedge T_A)$ is known
as the Nieh-Yan four-form \cite{nayf} and is actually an additional 
contribution to the ABJ-anomaly when the torsion is nonvanishing 
\cite{mielke}. However, as it is apparent from dimensional arguments, 
the Nieh-Yan four-form appears 
in the ABJ anomaly multiplied by the square of the regulator mass if a
Pauli-Villars-Gupta scheme, such as the one proposed here, is used i.e.
\begin{eqnarray}
\partial_\mu \langle J^\mu \rangle
&\propto& {\Lambda^2}*d(e^A \wedge T_A) \cr
\nonumber\\
&+& ({\rm usual} *Tr(F\wedge F)\, {\rm and} *Tr(G\wedge G)\, {\rm terms})
+ O(\Lambda^{-2}).
\nonumber\\
\end{eqnarray}
Recall that $\Lambda$ is the regulator mass scale. The additional contribution
to the ABJ anomaly when there is torsion diverges as the regulator mass 
is taken to infinity. Therefore the integral of the Nieh-Yan 
four-form will appear as a counterterm in the 
effective action of the Weyl theory. So
when the torsion is nonvanishing, there is an additional boundary
counterterm from the imaginary part of the Weyl action which corresponds to
 the boundary term in the Samuel-Jacobson-Smolin action and accounts for the
first term in Eq.(110). The other imaginary
term $i\int_M B_\mu J^\mu$ in the Weyl action generates fermion loops
from the expansion of $\langle J^\mu \rangle$ which gives rise to torsion
counterterms including the torsion-squared term
which is the second term of Eq.(110) in the Samuel-Jacobson-Smolin action.

\newpage
\bigskip
\section*{VIII. REMARKS}
\bigskip

We have compared the conventional prescription for the interaction of matter
and the four known forces with one which extends the 
chiral (Weyl) coupling to include gravity in four dimensions.
In this regard, when fermions and the standard model are incorporated 
to reproduce the observed phenomena,
the differences studied here can serve as tests of the
attempts to quantize gravity with (anti)-self-dual variables without
having to rely exclusively on predictions from the full quantum theory 
of gravity.  

%Whatever the ultimate theory may turn out to be, it is intriguing that 
%the Weyl nature of the interaction
%between matter and forces in four dimensions can be extended to 
%encompass gravitation 
%The departures from the conventional prescription are indeed subtle. 
At the low energy classical level of equations of motion, the 
alternatives are indistinguishable. As far as local symmetries
are concerned, there are no inconsistencies for the completely
chiral prescription since there are no local gauge, Lorentz or
diffeomorphism anomalies. Moreover, a regularization which preserves
these invariances and the Weyl nature of the theory exists.
In the chiral alternative, the Weyl nature supersedes Hermiticity
and CPT invariance is not automatic.
Off-shell, CPT is violated when the torsion component $B_\mu$ is 
nonvanishing. At the level of quantum field theory in curved spacetimes with 
nonvanishing torsion, there are detectable differences because the ABJ 
anomaly exists and because one-loop counterterms generated by the 
alternatives are different. These differences serve to characterize 
the chiral nature of the interaction of matter and the known
forces at an even more fundamental
level than current empirical data may suggest.

%
%
%It behooves us to reconsider our understanding and to be on 
%the lookout for these differences, for Nature may be Weyl at an even more 
%fundamental level than current empirical data may suggest.
%

\bigskip\bigskip\bigskip
\section*{ACKNOWLEDGMENTS}
\bigskip

The research for this work has been supported in part by 
the Natural Sciences and Engineering Research Council of Canada 
and the Department of Energy under Grant No. DE-FG05-92ER40709-A005.
C.S. would like to thank Lee Smolin for his encouragement 
and keen interest in this work.


\bigskip
\begin{thebibliography}{99}
\bigskip

\bibitem{ash}A.\ Ashtekar, Phys.\ Rev.\ Lett. 
{\bf 57}, 2244 (1986);
Phys.\ Rev. {\bf D36}, 1587(1986);  {\it Lectures on nonperturbative
canonical gravity}, (World Scientific, Singapore, 1991) and references
therein.

\bibitem{samuel} J.\ Samuel, Pram$\bar{\rm a}$na J.\ Phys. {\bf 28}, 
L429(1987);
Class.\ Quantum Grav. {\bf 5}, L123 (1988); T.\ Jacobson and L.\ Smolin,
Phys. Lett. {\bf B196}, 39 (1987), Class. Quantum Grav. {\bf 5}, 583 (1988).

\bibitem{soo} C. Soo, Phys. Rev. D {\bf 52}, 3484 (1995). 

\bibitem{cps} L. N. Chang and C. Soo, Phys. Rev. D {\bf 53}, 5682 (1996).

\bibitem{lee}
W. Pauli, {\it Exclusion Principle, Lorentz Group and Reflection
of Space-time and Charge}, in {\it Niels Bohr and the Development
of Physics} edited by W. Pauli, L. Rosenfeld and V. Weisskopf  
(McGraw-Hill, New York, 1955) and references to earlier works of
L\"{u}ders and Schwinger mentioned therein;
G. L\"{u}ders, Ann. Phys. {\bf 2}, 1 (1957); 
see also, for instance, Chapter 14 of {\it Particle Physics and 
Introduction to Field Theory} by T. D. Lee 
(Harwood Academic Publishers, 1981).

\bibitem{inv} L. N. Chang and C. Soo, Phys. Rev. D {\bf 55}, 2410 (1997).

\bibitem{ABJ}
S. L. Adler,
Phys. Rev. {\bf 177}, 2426 (1969);
J. S. Bell and R. Jackiw, Nuovo Cimento {\bf 60A}, 47 (1969);
W. A. Bardeen, Phys. Rev. {\bf 184}, 1848 (1969).

\bibitem{glash} See, for instance, H. Georgi and S. L. Glashow,
Phys. Rev. Lett. {\bf 32}, 438 (1974); H.\ Georgi, 
in {\it Particles and Fields - 1974}, 
edited by C. E. Carlson, (AIP Conf. Proc. No. 23, New York, 1975);
H.\ Fritzsch and P.\ Minkowski, Ann. Phys. {\bf 93}, 193 (1975).

\bibitem{ART} T. Jacobson, Class. Quantum Grav. {\bf 5}, L143 (1988);
A.\ Ashtekar, J.\ D.\ Romano and R.\ S.\ Tate, Phys. Rev. D {\bf
40}, 2572 (1989); H.\ Kodama, Int. J. Mod. Phys. {\bf D1}, 439 (1993).

\bibitem{Fujienergy}
K. Fujikawa, Phys. Rev. D{\bf 29} 285 (1984).

\bibitem{nay} H. T. Nieh and M. L. Yan, Ann. Phys. {\bf 138}, 237 (1982).

\bibitem{Pauli}
W. Pauli and F. Villars,
Rev. Mod. Phys. {\bf 21}, 434 (1949); S. N. Gupta, Proc. Phys. Soc.
{\bf A66}, 129 (1953).

\bibitem{FS} S. A. Frolov and A. A. Slavnov,
Phys. Lett. {\bf B296}, 159 (1992).

\bibitem{nar}
R. Narayanan and H. Neuberger,
Phys. Lett. {\bf B302}, 62 (1993);
Phys. Rev. Lett. {\bf 71} 3251 (1993);
Nucl. Phys. {\bf B412}, 574 (1994); {\bf B443}, 305 (1995).

\bibitem{Okuyama}
K. Okuyama and H. Suzuki, hep-th/9603062; 
Phys. Lett. {\bf B382}, 117 (1996).

\bibitem{gglash} H. Georgi and S. Glashow, Phys. Rev. D{\bf 6}, 429 (1972); 
D. J. Gross and R. Jackiw, {\it ibid.} {\bf 6}, 477 (1972); C. Bouchiat, J.
Illiopoulos, and P. Meyer, Phys. Lett. {\bf 38B}, 519 (1972).

\bibitem{nieh} H. T. Nieh, Phys. Rev. Lett. {\bf 53}, 2219 (1984);
L. Alvarez-Gaume and E. Witten, Nucl. Phys. {\bf B234}, 269 
(1984); S. Yajima and T. Kimura, Phys. Lett. {\bf B173}, 154 (1986).

\bibitem{Chang} L. N. Chang and H. T. Nieh, Phys. Rev. Lett. {\bf 53}, 
21 (1984); also the second reference in \cite{nieh}.

\bibitem{fuji}
K. Fujikawa,
Phys. Rev. D{\bf 25} 2584 (1982); D{\bf 21}, 2848 (1980); 
D {\bf 22} 1499(E) (1980); D {\bf 23}, 2262 (1981);
Phys. Rev. Lett. {\bf 42} 1195 (1979); {\it ibid.} {\bf 44}, 1733 (1980).

\bibitem{nayf} H. T. Nieh and M. L. Yan, J. Math. Phys. {\bf 23}, 373 (1982).

\bibitem{mielke} Two recent works on the relation of the Nieh-Yan
four-form to the ABJ-anomaly are
O. Chandia and J. Zanelli, hep-th/9702025; Y. N. Obukhov, E. W. Mielke,
J. Budczies and F. W. Hehl, gr-qc/9702011.

%\bibitem{Birrell} See, for instance, {\it Quantum fields in curved spaces} by
%N. D. Birrell and P. C. W. Davies (Cambridge University Press, 1982).

\end{thebibliography}
\end{document}